\documentclass[a4paper]{jpconf}
\usepackage{graphicx}
\usepackage{bm}
\usepackage{epsfig}
\usepackage{graphics}
\usepackage{amsmath}
\usepackage{amssymb}
\usepackage{latexsym}
\usepackage{slashed,cite}
\usepackage{upgreek}
\usepackage[l]{floatflt}

\renewenvironment{subequations}{%
\refstepcounter{equation}%
\setcounter{parentequation}{\value{equation}}%
  \setcounter{equation}{0}
  \ignorespaces
}{%
  \setcounter{equation}{\value{parentequation}}%
  \ignorespacesafterend
}

\newcommand{\beqs}{\begin{subequations}}
\newcommand{\eeqs}{\end{subequations}}
\newcommand{\eec}{\end{center}}
\newcommand{\bec}{\begin{center}}
\newcommand{\eem}{\end{matrix}}
\newcommand{\bem}{\begin{matrix}}
\newcommand{\Eref}[1]{Eq.~(\ref{#1})}
\renewcommand{\Sref}[1]{Sec.~\ref{#1}}
\renewcommand{\Fref}[1]{Fig.~\ref{#1}}
\renewcommand{\Tref}[1]{Table~\ref{#1}}
\newcommand{\cref}[1]{Ref.~\cite{#1}}

\newcommand\eqs[2]{Eqs.~(\ref{#1}) and (\ref{#2})}

\newcommand{\sEref}[2]{Eq.~(\ref{#1}{\small\sf {#2}})}

\newcommand{\fr}{\ensuremath{f_{1}}}
\newcommand{\frr}{\ensuremath{f_{2}}}
\newcommand{\fk}{\ensuremath{f_p}}
\newcommand{\fkk}{\ensuremath{N_p}}

\newcommand{\ks}{\ensuremath{k_\star}}

\newcommand{\eeq}{\end{equation}}
\newcommand{\beq}{\begin{equation}}
\newcommand{\ba}{\begin{array}}
\newcommand{\ea}{\end{array}}
\newcommand{\bea}{\begin{eqnarray}}
\newcommand{\eea}{\end{eqnarray}}
\newcommand{\ftn}{\footnotesize}

\renewcommand{\Re}{{\mbox{\sf\small Re}}}

\newcommand{\hepph}[1]{{\ftn \tt hep-ph/#1}}

\newcommand{\astroph}[1]{{\ftn \tt astro-ph/#1}}
\newcommand{\arxiv}[1]{{\ftn\tt  arXiv:#1}}

\newcommand\vev[1]{\langle {#1} \rangle}
\newcommand\vevi[1]{\langle {#1} \rangle_{\rm I}}
\def\lf{\left(}
\def\rg{\right)}

\newcommand{\Vhi}{\ensuremath{V_{\rm I}}}
\newcommand{\Hhi}{\ensuremath{H_{\rm I}}}
\newcommand{\as}{\ensuremath{\alpha_{\rm s}}}
\newcommand{\wrh}{\ensuremath{w_{\rm rh}}}

\newcommand{\Dex}{\ensuremath{\Delta_{\star}}}

\newcommand{\Ve}{\ensuremath{V}}

\newcommand{\Qef}{\ensuremath{\Lambda_{\rm UV}}}

\def\bbet{{\bar\beta}}
\def\al{{\alpha}}
\def\bt{{\beta}}

\def\n{\bar{n}}

\newcommand{\Trh}{\ensuremath{T_{\rm rh}}}

\newcommand{\ld}{\ensuremath{\lambda}}

\newcommand{\kp}{\ensuremath{\kappa}}
\newcommand{\se}{\ensuremath{\widehat\phi}}
\newcommand{\sex}{\ensuremath{\widehat{\phi}_*}}
\newcommand{\sgx}{\ensuremath{\phi_\star}}
\newcommand{\lda}{\ensuremath{\lambda_1}}
\newcommand{\ldb}{\ensuremath{\lambda_2}}

\newcommand{\ldd}{\ensuremath{\lambda_4}}

\newcommand{\sef}{\ensuremath{\widehat{\phi}_{\rm f}}}

\newcommand{\eph}{\ensuremath{\widehat \epsilon}}
\newcommand{\ith}{\ensuremath{\widehat \eta}}
\newcommand{\mP}{\ensuremath{m_{\rm P}}}
\def\Ka{K\"{a}hler potential}

\newcommand{\sg}{\ensuremath{\phi}}

\renewcommand{\sigma}{\ensuremath{\phi}}

\newcommand{\sgf}{\ensuremath{\phi_{\rm f}}}

\newcommand{\GeV}{{\mbox{\rm GeV}}}
\newcommand{\what}{\ensuremath{\widehat}}
\newcommand{\Khi}{\ensuremath{K}}

\newcommand{\Vhio}{\ensuremath{V_{\rm I0}}}

\newcommand{\diag}{\mbox{\sf\small diag}}

\newcommand{\Mgut}{\ensuremath{M_{\rm GUT}}}
\newcommand{\Ggut}{\ensuremath{G_{\rm GUT}}}
\newcommand{\Gsm}{\ensuremath{G_{\rm SM}}}

\newcommand{\mbl}{\ensuremath{M_{BL}}}

\newcommand{\mgut}{\ensuremath{M_{\rm GUT}}}

\newcommand{\ldu}{\ensuremath{\uplambda}}
\renewcommand{\ns}{\ensuremath{n_{\rm s}}}
\newcommand{\As}{\ensuremath{A_{\rm s}}}
\newcommand{\Ns}{\ensuremath{N_{\star}}}

\def\th{{\theta}}
\def\thb{{\bar\theta}}
\def\thn{{\theta_{\Phi}}}

\newcommand{\phc}{\ensuremath{\Phi}}
\newcommand{\phcb}{\ensuremath{\bar\Phi}}

\newcommand{\sgm}{\ensuremath{\phi_{\rm mx}}}
\newcommand{\sgn}{\ensuremath{\phi_{\rm mn}}}

\newcommand{\mma}{\ensuremath{M_1}}
\newcommand{\mmb}{\ensuremath{M_2}}
\newcommand{\mmc}{\ensuremath{M_4}}

\newcommand{\ts}{\ensuremath{\delta_{42}}}
\newcommand{\rs}{\ensuremath{\delta_{21}}}
\newcommand{\rss}{\ensuremath{r_{21}}}
\newcommand{\rrs}{\ensuremath{r_{12}}}
\newcommand{\tss}{\ensuremath{r_{42}}}
\newcommand{\tts}{\ensuremath{r_{24}}}

\renewcommand{\Dex}{\ensuremath{\Delta_{\rm \star}}}

\newcommand{\kbaa}{\ensuremath{{K_{2(11)^2}}}}

\newcommand{\tkbaa}{\ensuremath{{\widetilde K_{2(11)^2}}}}

\newcommand{\kas}{\ensuremath{K_{1\rm s}}}
\newcommand{\tkas}{\ensuremath{\widetilde K_{1\rm s}}}
\newcommand{\kbas}{\ensuremath{K_{21\rm s}}}
\newcommand{\tkbas}{\ensuremath{\widetilde K_{21\rm s}}}

\newcommand{\kb}{\ensuremath{K_{2}}}
\newcommand{\kaa}{\ensuremath{K_{(11)^2}}}
\newcommand{\tkaa}{\ensuremath{\widetilde K_{(11)^2}}}

\newcommand{\dci}{{\small\sf $\delta$EM}}
\newcommand{\ca}{{\small\sf EM2}}
\newcommand{\cb}{{\small\sf EM4}}
\newcommand{\dhi}{{\small\sf $\delta$TM}}
\newcommand{\ha}{{\small\sf TM4}}
\newcommand{\hb}{{\small\sf TM8}}

\newcommand{\tmd}{{TMI}}
\newcommand{\emd}{{EMI}}

\begin{document}

\title{Formulating E- \& T-Model Inflation in Supergravity}

\author{C Pallis}

\address{Laboratory of Physics, Faculty of
Engineering, \\ Aristotle University of Thessaloniki,
Thessaloniki\\ GR-541 24, Greece}

\ead{kpallis@gen.auth.gr}

\date{\today}

\begin{abstract} We present novel realizations of E- and T-model inflation within
Supergravity which are largely associated with the existence of a
pole of order one and two respectively in the kinetic term of the
inflaton superfield. This pole arises due to the selected
logarithmic Kahler potentials $K$, which parameterize hyperbolic
manifolds with scalar curvature related to the coefficient
$(-N)<0$ of a logarithmic term. The associated superpotential $W$
exhibits the same $R$ charge with the inflaton-accompanying
superfield and includes all the allowed terms. The role of the
inflaton can be played by a gauge singlet or non-singlet
superfield. Models with one logarithmic term in $K$ for the
inflaton, require $N=2$, some tuning -- of the order of $10^{-5}$
-- between the terms of $W$ and predict a tensor-to-scalar ratio
$r$ at the level of $0.001$. The tuning can be totally eluded for
more structured $K$'s, with $N$ values increasing with $r$ and
spectral index close or even equal to its present central
observational value.

\end{abstract}

\section{Introduction}\label{intro}

It is well-known \cite{terada,sor,epole} that the presence of a
pole in the kinetic term of the inflaton gives rise to
inflationary models collectively named $\alpha$-attractors
\cite{alinde, eno5}. Confining ourselves to poles of order one and
two, the lagrangian of the homogenous inflaton field $\sg=\sg(t)$
can be written as
\beq \label{action1} {\cal  L} = \sqrt{-\mathfrak{g}}
\left(\frac{\fkk}{2\fk^2} \dot\sg^2-
\Vhi(\sg)\right)~~\mbox{with}~~\fk=1-\sg^p,~p=1,2~~\mbox{and}~~\fkk>0.\eeq
Also, $\mathfrak{g}$ is the determinant of the background
Friedmann-Robertson-Walker metric $g^{\mu\nu}$ with signature
$(+,-,-,-)$, dot stands for derivation \emph{with respect to}
({\sf\small w.r.t}) the cosmic time and $V_{\rm I}=V_{\rm I}(\sg)$
is a polynomial function of $\sg$. Expressing the canonically
normalized field, $\se$, in terms of $\sg$ as follows
\beq \label{VJe}
\frac{d\se}{d\phi}=J=\frac{\sqrt{\fkk}}{\fk}~~\Rightarrow~~\sg=
\begin{cases}
1-e^{-\se/\sqrt{N_1}} &\mbox{for}~~p=1,\\
\tanh{\lf\frac{\se}{\sqrt{N_2}}\rg} &\mbox{for}~~p=2\,,
\end{cases}
\eeq
we easily infer that  $V_{\rm I}(\se)$ develops a plateau for
$\se\gg1$, and so it becomes suitable for driving inflation of
chaotic type. It is called \emph{E-Model Inflation} ({\sf\small
EMI}) \cite{alinde, linde21} (or $\alpha$-Starobinsky model
\cite{ellis21}) and \emph{T-Model Inflation} ({\sf\small \tmd})
\cite{tmodel, linde21} for $p=1$ and $2$ respectively.

In this talk, based on \cref{sor, epole}, we present novel
realizations of \emph{E and/or T-Model Inflation} ({\sf\small
ETI}) in the context of \emph{Supergravity} ({\sf\small SUGRA}).
Namely, in \Sref{fhim} we describe our general strategy in
establishing those models and then -- in \Sref{fhi3} -- we specify
three models (\dci, \ca, \cb) employing a gauge singlet inflaton
and three (\dhi, \ha, \hb) using a gauge non-singlet inflaton. We
end up with our numerical results in \Sref{num} and our
conclusions in \Sref{con}.

Throughout the text, the subscript $,\chi$ denotes derivation
w.r.t the field $\chi$, charge conjugation is denoted by a star
($^*$) and we use units where the reduced Planck scale $\mP =
2.44\cdot 10^{18}~\GeV$ is set equal to unity.

\section{Strategy of the SUGRA Embedding}\label{fhim}

We start our investigation presenting the basic formulation of
scalar theory within SUGRA in \Sref{sugra1} and then -- in
\Sref{sugra2} and \ref{sugra3} -- we outline our methodology in
constructing viable ETI.

\subsection{General Formalism} \label{sugra1}

The part of the SUGRA lagrangian including the (complex) scalar
fields $z^\al$ can be written as
\beq\label{Saction1} {\cal  L} = \sqrt{-\mathfrak{g}} \lf
K_{\al\bbet} D_\mu z^\al D^\mu z^{*\bbet}-V_{\rm SUGRA}\rg, \eeq
where the kinetic mixing is controlled by the K\"ahler potential
$K$ and the relevant metric defined as
\beqs \beq \label{kddef} K_{\al\bbet}={\Khi_{,z^\al
z^{*\bbet}}}>0\>\>\>\mbox{with}\>\>\>K^{\bbet\al}K_{\al\bar
\gamma}=\delta^\bbet_{\bar \gamma}.\eeq
Also, the covariant derivatives for the scalar fields $z^\al$ are
given by
\beq D_\mu z^\al=\partial_\mu z^\al+ig A^{\rm a}_\mu T^{\rm
a}_{\al\bt} z^\bt\eeq\eeqs
with $A^{\rm a}_\mu$ being the vector gauge fields, $g$ the
(unified) gauge coupling constant and $T^{\rm a}$ with ${\rm
a}=1,...,\mbox{\sf\small dim}\Ggut$ the generators of a
\emph{Grand Unified Theory} ({\sf \ftn GUT}) gauge group $\Ggut$.
Here and henceforth, the scalar components of the various
superfields are denoted by the same superfield symbol.

The SUGRA scalar potential, $V_{\rm SUGRA}$, is given in terms of
$K$, and the superpotential, $W$, by
\beq V_{\rm SUGRA}=\Ve_{\rm F}+ \Ve_{\rm D}\>\>\>\mbox{with}\>\>\>
\Ve_{\rm F}=e^{\Khi}\left(K^{\al\bbet}D_\al W D^*_\bbet
W^*-3{\vert W\vert^2}\right) \>\>\>\mbox{and}\>\>\>\Ve_{\rm D}=
g^2 \mbox{$\sum_{\rm a}$} {\rm D}_{\rm a}^2/2. \label{Vsugra} \eeq
The K\"ahler covariant derivative and the D-terms read
\beq \label{Kinv} D_\al W=W_{,z^\al}
+K_{,z^\al}W\>\>\>\mbox{and}\>\>\>{\rm D}_{\rm a}= z_\al\lf T_{\rm
a}\rg^\al_\bt K^\bt\>\>\>\mbox{with}\>\>\>
K^{\al}={\Khi_{,z^\al}}\,.\eeq
Therefore, ETI introduced in \Sref{intro} can be supersymmetrized,
if we select conveniently the functions $K$ and $W$ so that
\Eref{Saction1} reproduces \Eref{action1}.

\subsection{Inflaton's Sector}\label{sugra2}

We concentrate on ETI driven by $V_{\rm F}$ assuring that $V_{\rm
D}=0$ during it. This condition may be attained identifying the
inflaton either with (the radial part of) a gauge singlet
superfield $z^2:=\Phi$ or with  the radial part of a conjugate
pair of Higgs superfields, $z^2:=\Phi$ and $z^3:=\bar\Phi$.

To achieve a kinetic term in \Eref{Saction1} similar to that in
\Eref{action1} for $p=1$ and $2$, we need to establish suitable
$K$'s so that (during ETI)
\beq \vevi{K}= -N
\ln\fk~~~\mbox{and}~~~\vevi{K_{\al\bbet}}=N/\fk^2\eeq
with $N$ related to $N_p$. However, from the F-term contribution
to \Eref{Vsugra}, we remark that $K$ affects -- besides the
kinetic mixing -- $V_{\rm SUGRA}$, which, in turn, depends on $W$
too. Therefore, $\fk$ is generically expected to emerge also in
the denominator of $V_{\rm SUGRA}$ making difficult the
establishment of an inflationary era. This problem can be
addressed \cite{sor, epole} either by tuning the terms of $W$ so
that the pole is removed from $V_{\rm SUGRA}$ thanks to
cancellations or adopting a more structured $K$ which yields the
desired kinetic terms in \Eref{action1} but lets $V_{\rm SUGRA}$
immune.

\subsection{Stabilizer's Sector}\label{sugra3}

We reserved $\al=1$ for a gauge singlet superfield, $z^1=S$ called
stabilizer or goldstino, which assists \cite{rube} us to formulate
ETI of chaotic type in SUGRA. We assume that $S$ and $W$ are
equally charged under a global $R$ symmetry and so it appears
linearly in $W$ multiplying its other terms. Also, it generates
for $\vevi{S}=0$ the inflationary potential via the only term of
$V_{\rm SUGRA}$ which remains alive
\beq \Vhi=\vevi{V_{\rm F}}= \vevi{e^{K}K^{SS^*}|W_{,S}|^2},
\label{Vhio}\eeq
where the symbol ``$\vevi{Q}$" denotes the value of a quantity $Q$
during ETI. Such an adjustment assures the boundedness of $\Vhi$,
whereas $S$ can be stabilized at $\vevi{S}=0$, if we select
\cite{su11}
\beq \label{K2}
K_2=N_S\ln\lf1+|S|^2/N_S\rg~~\Rightarrow~~\vevi{K_2^{SS^*}}=1~~\mbox{with
$0<N_S<6$}.\eeq $K_2$ parameterizes the compact manifold
$SU(2)/U(1)$. Note that for $\vevi{S}=0$, $S$ is canonically
normalized and so we do not mention it again henceforth.

\section{Inflationary Settings}\label{fhi3}

We here explain the construction of the EMI and TMI in
\Sref{fhi33} and \ref{fhi1} respectively.

\subsection{E-Model Inflation}\label{fhi33}

This setting is realized in presence of two gauge singlet
superfields $S$ and $\Phi$. The relevant SUGRA setup is presented
in \Sref{fhi30} whereas the resulting models are derived in
\Sref{fhi31}.

\subsubsection{SUGRA Set-up}\label{fhi30}

We adopt the most general renormalizable $W$ consistent with the
$R$ symmetry mentioned in \Sref{sugra3}, i.e.,
\beq W= S\lf \lda \phc+\ldb\phc^2-M^2\rg \label{whi} \eeq
where $\lda, \ldb$ and $M$ are free parameters. As regards $K$,
this includes, besides $K_2$ in \Eref{K2}, one of the following
$K$'s:
\beq \kas=-N\ln\left(1-\phc/2-\phc^*/2\right)~~~\mbox{or}~~~
\tkas=-N\ln\frac{(1-\phc/2-\phc^*/2)}{(1-\phc)^{1/2}(1-\phc^*)^{1/2}},\label{tkas}\eeq
with $\Re(\phc)<1$ and $N>0$.  For the considered versions of EMI,
\dci\ and \ca -- \cb, the total $K$ respectively is
\beq {\sf\small (a)}~\kbas=\kb+\kas~~\mbox{and}~~{\sf\small
(b)}~\tkbas=\kb+\tkas. \label{kbas}\eeq The elimination of pole in
$\Vhi$ for \dci\ can be achieved if we set \beq
N=2~~~\mbox{and}~~~\rss=-\ldb/\lda\simeq
1+\rs\,~~\mbox{with}~~~\rs\sim0~~~\mbox{and}~~M\ll1.\label{kbas1}
\eeq
No denominator exists in $\Vhi$ for \ca\ and \cb\ and so $N$,
$\lda$, $\ldb$ and $M$ are free parameters.

\subsubsection{Inflationary Configuration}\label{fhi31}

The inflationary potential can be derived from \Eref{Vhio}
specifying the inflationary trajectory as follows
\beq \vevi{S}=0\>\>\>\mbox{and}\>\>\>\vevi{\th}:=\mbox{\sf\small
arg}\vevi{\Phi}=0. \label{inftr3}\eeq
Inserting the quantities above into \Eref{Vhio} and taking into
account \Eref{K2} and
\beq \label{eK} \vevi{e^{K}}=\begin{cases}
\fr^{-N}&\mbox{for}\>\>\>K=\kbas,\\
1& \mbox{for}\>\>\>K=\tkbas,
\end{cases}\eeq
we arrive at the following master equation
\beq\Vhi=\ld^2 \begin{cases}
%\begin{array}{rl}
{\lf  \sg-\rss\sg^2-\mma^2\rg^2}/{\fr^{N}}&\mbox{for \dci},\\
{\lf \sg-\rss\sg^2-\mma^2\rg^2}&\mbox{for \ca},\\
{\lf \sg^2-\rrs\sg-\mmb^2\rg^2}&\mbox{for
\cb},\end{cases}\label{Vmab}\eeq
where $\sg=\Re(\Phi)$, $r_{ij}=-\ld_i/\ld_j$ with $i,j=1,2$ and
$\ld$ and $M_i$ are identified as follows
\beq\ld= \begin{cases}
%\begin{array}{rl}
\lda~~\mbox{and}~~\mma=M/\sqrt{\lda}&\mbox{for~\dci\ and \ca},\\
\ldb~~\mbox{and}~~\mmb=M/\sqrt{\ldb}&\mbox{for~\cb}.\end{cases}\label{ldab}\eeq
As advertised in \Sref{fhi30}, the pole in $\fr$ is presumably
present in $\Vhi$ for \dci, but it disappears for \ca\ and \cb.
The arrangement of \Eref{kbas1}, though, renders the pole harmless
for \dci.

We introduce the canonically normalized fields, $\se$ and $\what
\th$, as follows
\beq \label{K3} \vevi{K_{\Phi\Phi^*}}|\dot \Phi|^2
\simeq\frac12\lf\dot{\what\phi}^{2}+\dot{\what
\th}^{2}\rg~~\Rightarrow~~\frac{d\se}{d\sg}=J={\sqrt{N/2}\over\fr}~~\mbox{and}~~
\widehat{\theta}\simeq
J\sg\theta~~~\mbox{with}~~~\vevi{K_{\Phi\Phi^*}}=\frac{N}{4f_1^2}.
\eeq
We see that the relation between $\sg$ and $\se$ is identical with
\Eref{VJe} for $p=1$, if we do the replacement $N_1=N/2$.

\renewcommand{\arraystretch}{1.4}
\begin{table}[!t]
\bec\begin{tabular}{cccll}\br {\sc Fields}&{\sc Eigen-}&
\multicolumn{3}{c}{\sc Masses Squared}\\\cline{3-5} &{\sc states}&
& {$K=\kbas$}&{$K=\tkbas$}\cr \hline\hline $1$ real scalar
&$\widehat \theta$ & $\widehat m^2_{\theta}$&
\multicolumn{2}{c}{$6H_{\rm I}^2$}\cr
$2$ real scalars &$\what s_1,~\what s_2$ & $\what m^2_{
s}$&\multicolumn{2}{c}{$6H_{\rm I}^2/N_S$}\cr\mr
$2$ Weyl spinors & ${(\what{\psi}_{\Phi}\pm
\what{\psi}_{S})/\sqrt{2}}$& $\what m^2_{ \psi\pm}$&
\multicolumn{2}{c}{$6n(1-\sg)^2H_{\rm I}^2/N\sg^2$}\cr\br
\end{tabular}\eec
\caption{\sl Mass spectrum of our EMI along the trajectory of
\Eref{inftr3} -- we take $n=1$ for \textsf{$\delta$\small EM} and
\textsf{\small EM2} whereas $n=2$ for \textsf{\small
EM4}.}\label{tab3}
\end{table}\renewcommand{\arraystretch}{1}

To check the stability of $V_{\rm SUGRA}$ in \Eref{Vsugra} along
the trajectory in \Eref{inftr3} w.r.t the fluctuations of
$z^\alpha$'s, we construct the mass spectrum of the theory. Our
results are summarized in \Tref{tab3}. Taking the limit
$\rs=\mma=0$ for \dci, $\rss=\mma=0$ for \ca\ and $\rrs=\mmb=0$
for \cb, we find the expressions of the masses squared $\what
m^2_{\chi^\al}$ (with $\chi^\al=\th$ and $s$) arranged in
\Tref{tab3}. We there display the masses $\what m^2_{\psi^\pm}$ of
the corresponding fermions too -- we define
$\what\psi_{\Phi}=J\psi_{\Phi}$ where $\psi_\Phi$ and $\psi_S$ are
the Weyl spinors associated with $S$ and $\Phi$ respectively. We
notice that the relevant expressions can take a unified form for
all models -- recall that we use $N=2$ in \dci\ -- and approach,
close to $\sg=\sgx\simeq1$, rather well the quite lengthy, exact
ones employed in our numerical computation. From them we can
appreciate the role of $N_S<6$ in retaining positive $\what
m^2_{s}$. Also, we confirm that $\what
m^2_{\chi^\al}\gg\Hhi^2\simeq\Vhio/3$ for $\sgf\leq\sg\leq\sgx$.

\subsection{T-Model Inflation}\label{fhi1}

Although TMI is mostly realized by a gauge-singlet inflaton
\cite{alinde}, we here propose an alternative scheme where the
inflaton is included in a conjugate pair of Higgs superfields,
$\bar\Phi$ and $\Phi$ with charges $B-L=-1$ and $1$ respectively.
These fields break the GUT symmetry $\Ggut=\Gsm\times U(1)_{B-L}$
down to the \emph{Standard Model} ({\small\sf SM}) gauge group
$\Gsm$ through their \emph{vacuum expectation values} ({\small\sf
v.e.vs}). We below outline the SUGRA setting in \Sref{fhi10} and
its inflationary outcome in \Sref{fhi11}.

\subsubsection{SUGRA Set-up}\label{fhi10}

Consistently with the imposed symmetries we adopt the following
$W$
\beq W= S\lf \frac12\ldb\bar\Phi\Phi+\ldd
(\bar\Phi\Phi)^2-\frac14M^2\rg,\label{whih} \eeq
where $\ldb, \ldd$ and $M$ are free parameters. This type of $W$
is well-known from the models of F-term hybrid inflation -- for
reviews see, e.g., \cref{fhi}. The invariance of $K$ under $\Ggut$
enforces us to introduce a pole of order two in the $\phcb-\phc$
kinetic terms. One possible option -- for another equivalent one
see \cref{sor} -- is
\beq\kaa=-\frac{N}{2}\ln\left(1-2|\phc|^2\right)\left(1-2|\phcb|^2\right)\>\>\>\mbox{or}\>\>\>
\tkaa=-\frac{N}{2}\ln\frac{\left(1-2|\phc|^2\right)\left(1-2|\phcb|^2\right)}
{(1-2\phcb\phc)^{1/2}(1-2\phcb^*\phc^*)^{1/2}},\label{tkba}\eeq
which parameterizes the manifold ${\cal
M}_{(11)^2}=SU(1,1)/U(1)_\Phi \times SU(1,1)/U(1)_{\bar\Phi}$
\cite{sor} with scalar curvature ${\cal R}_{(11)^2}=-{8}/{N}$.
From the selected above $W$ and $K$'s, we obtain the models \dhi\
and \ha\ -- \hb, which employ the following total $K$'s
correspondingly
\beq {\sf\small (a)}~\kbaa=\kb+\kaa~~\mbox{and}~~{\sf\small
(b)}~\tkbaa=\kb+\tkaa. \label{kbba} \eeq
Within \dhi, an elimination of the singular denominator appearing
in $\Vhi$ is obtained setting
\beq N=2~~~\mbox{and}~~~\tss=-\ldd/\ldb\simeq
1+\ts\,~~\mbox{with}~~~\ts\sim0~~~\mbox{and}~~~M\ll1.\label{kbba1}
\eeq
On the other hand, no singularity exists in $\Vhi$, within \ha\
and \hb\ and so $N$, $\ldb$, $\ldd$ and $M$ remain free
parameters.

\subsubsection{Inflationary Configuration}\label{fhi11}

Adopting for the relevant fields the parameterization  \beq
\Phi=\phi e^{i\theta}\cos\theta_\Phi\>\>\>
\mbox{and}\>\>\>\bar\Phi=\phi e^{i\thb}\sin\theta_\Phi\>\>\>
\mbox{with}\>\>\>0\leq\thn\leq{\pi}/{2}~~~\mbox{and}~~~S= \lf{s
+i\bar s}\rg/{\sqrt{2}},\eeq
we can easily verify that a D-flat direction is \beq
\vevi{\theta}=\vevi{\thb}=0,\>\vevi{\thn}={\pi/4}\>\>\>\mbox{and}\>\>\>\vevi{S}=0,\label{inftr}\eeq
which can be qualified as inflationary path. Regarding the
exponential prefactor of $V_{\rm F}$ in \Eref{Vsugra} we obtain
\beq \label{eKh} \vevi{e^{K}}=\begin{cases}
\frr^{-N}&\mbox{for}\>\>\>K=\kaa,\\
1& \mbox{for}\>\>\>K=\tkaa,\end{cases}\eeq
Substituting it and \eqs{K2}{whih} into \Eref{Vhio}, the
inflationary potential $\Vhi$ takes its master form
\beq\Vhi=\frac{\ld^2}{16} \begin{cases}
%\begin{array}{rl}
{\lf  \sg^2-\tss\sg^4-\mmb^2\rg^2}/{\frr^{N}}&\mbox{for \dhi},\\
{\lf \sg^2-\tss\sg^4-\mmb^2\rg^2}&\mbox{for \ha},\\
{\lf \sg^4-\tts\sg^2-\mmc^2\rg^2}&\mbox{for
\hb},\end{cases}\label{Vmabh}\eeq
where $r_{ij}=-\ld_i/\ld_j$ with $i,j=1,2$ and $\ld$ and $M_i$ are
identified as follows
\beq\ld= \begin{cases}
%\begin{array}{rl}
\ldb~~\mbox{and}~~\mmb=M/\sqrt{\ldb}&\mbox{for~\dhi\ and \ha},\\
\ldd~~\mbox{and}~~\mmc=M/\sqrt{\ldd}&\mbox{for~\hb}.\end{cases}\label{ldbd}\eeq
From \Eref{Vmabh}, we infer that  the pole in $\frr$ is presumably
present in $\Vhi$ of \dhi\ but it disappears in $\Vhi$ of \ha\ and
\hb\ and so no $N$ dependence in $\Vhi$ arises. The elimination of
the pole in the regime of \Eref{kbba1} lets open the realization
of \dhi, though.

\renewcommand{\arraystretch}{1.4}
\begin{table}[!t]
\begin{center}
\lineup\begin{tabular}{cccll}\br {\sc Fields}&{\sc Eigen-}&
\multicolumn{3}{c}{\sc Masses Squared}\\\cline{3-5} &{\sc states}&
&
\multicolumn{1}{c}{$K=\kbaa$}&\multicolumn{1}{c}{$K=\tkbaa$}\cr\mr
%\hspace*{2.cm}
%
2 real&$\widehat\theta_{+}$&$m_{\widehat\theta+}^2$&
\multicolumn{2}{c}{$3\Hhi^2$}\cr
scalars&$\widehat \theta_\Phi$ &$\widehat m_{
\theta_\Phi}^2$&\multicolumn{2}{c}{$M^2_{BL}+6\Hhi^2(1+4/N-2/N\sg^2-2\sg^2/N)$}
\cr\mr
1 complex&$s, {\bar{s}}$ &$ \widehat m_{
s}^2$&\multicolumn{1}{c}{$6\Hhi^2(1/N_S-8(1-\sg^2)/N+N\sg^2/2$}&{$6\Hhi^2(1/N_S-4/N$}\cr
scalar&&&\multicolumn{1}{c}{$+2(1-2\sg^2)+8\sg^2/N)$}&\multicolumn{1}{c}{$+2/N\sg^2+2\sg^2/N)$}\cr\mr
1 gauge boson &{$A_{BL}$}&{$M_{BL}^2$}&
\multicolumn{2}{c}{$2Ng^2\sg^2/\frr^2$}\cr\mr
$4$ Weyl  & $\what \psi_\pm$ & $\what m^2_{ \psi\pm}$&
\multicolumn{2}{c}{${12\frr^2\Hhi^2/N^2\sg^2}$}\cr
spinors&$\ldu_{BL},
\widehat\psi_{\Phi-}$&$M_{BL}^2$&\multicolumn{2}{c}{$2Ng^2\sg^2/\frr^2$}\cr
\br
\end{tabular}\end{center}
\caption{\sl\small Mass spectrum for TMI along the inflationary
trajectory of \Eref{inftr}. }\label{tab1}
\end{table}\renewcommand{\arraystretch}{1.}

The canonically normalized (hatted) fields of the $\phcb-\phc$
system during TMI, are defined as follows
\beqs\beq \vevi{K_{\al\bbet}}\dot z^\al \dot z^{*\bbet} \simeq
\frac12\lf\dot{\widehat \sg}^2+\dot{\widehat
\th}_+^2+\dot{\widehat \th}_-^2+\dot{\widehat
\th}_\Phi^2\rg~~~\mbox{for}~~~\al=2,3. \label{kzzn}\eeq
where, via \Eref{tkba}, we find
\beq \lf \vevi{K_{\al\bbet}}\rg=
\vevi{M_{\phc\phcb}}~~\mbox{with}~~
\vevi{M_{\phc\phcb}}=\kp\,\diag(1,1)\>\>\mbox{and}~~
\kp={N}/{\frr^{2}}.\eeq\eeqs
Inserting the expressions above in \Eref{kzzn} we obtain the
hatted fields
\beq {d\se}/{d\sg}=J={\sqrt{2N}/\frr},~~\widehat{\theta}_\pm
={\sqrt{\kp}}\sg\theta_\pm\>\>\mbox{and}~~\widehat \theta_\Phi =
\sqrt{2\kp}\sg\lf\theta_\Phi-{\pi}/{4}\rg,\eeq
where $\th_{\pm}=\lf\bar\th\pm\th\rg/\sqrt{2}$. From the first
equation above we conclude that \Eref{VJe} for $p=2$ is reproduced
for $N_2=2N$.

We can also verify that the direction of \Eref{inftr} is stable
w.r.t the fluctuations of the non-inflaton fields. Approximate,
quite precise though, expressions for $\sg=\sgx\sim1$ are arranged
in \Tref{tab1}. We confine ourselves to the limits $\ts=\mmb=0$
for \dhi, $\tss=\mmb=0$ for \ha\ and $\tts=\mmc=0$ for \hb. As in
the case of the spectrum in \Tref{tab3}, $N_S<6$ plays a crucial
role in retaining positive and heavy enough $\what m^2_{s}$. Here,
however, we also display  the masses, $M_{BL}$, of the gauge boson
$A_{BL}$ (which signals the fact that $U(1)_{B-L}$ is broken
during TMI) and the masses of the corresponding fermions. It is
also evident that $A_{BL}$ becomes massive absorbing the massless
Goldstone boson associated with $\what\th_-$.

\section{Inflationary Observables}\label{num}

We here constrain the parameters of our \emd\ and \tmd\ -- in
\Sref{num1} and \ref{num2} respectively -- taking into account a
number of observational and theoretical requirements described in
\Sref{obs}.

\subsection{Constraints} \label{obs}

We impose to ETI the following observational and theoretical
requirements:

\subparagraph{\bf (a)} The number of e-foldings $\Ns$ that the
scale $\ks=0.05/{\rm Mpc}$ experiences during ETI must be enough
for the resolution of the  problems of standard Big Bang, i.e.,
\cite{plcp}
\begin{equation} \label{Nhi}  \Ns=\int_{\sef}^{\sex}
d\se\frac{\Vhi}{\Ve_{\rm I,\se}}\simeq61.3+\frac{1-3w_{\rm
rh}}{12(1+w_{\rm rh})}\ln\frac{\pi^2g_{\rm
rh}\Trh^4}{30\Vhi(\sgf)}+\frac14\ln{\Vhi(\sgx)^2\over g_{\rm
rh}^{1/3}\Vhi(\sgf)},\eeq
where $\sex$ is the value of $\se$ when $\ks$ crosses the
inflationary horizon whereas $\se_{\rm f}$ is the value of $\se$
at the end of ETI, which can be found, in the slow-roll
approximation, from the condition
\beqs\beq\mbox{\sf\small max}\{\epsilon(\sg_{\rm
f}),|\eta(\sg_{\rm f})|\}=1,\>\>\>~\mbox{where}\>\>\>\epsilon=
{1\over2}\left(\frac{\Ve_{\rm I,\se}}{\Ve_{\rm
I}}\right)^2\>\>\>\mbox{and}\>\>\>\eta= \frac{\Ve_{\rm
I,\se\se}}{\Ve_{\rm I}}\,.\label{sr}\eeq
Also we assume that ETI is followed in turn by an oscillatory
phase with mean equation-of-state parameter $w_{\rm rh}$,
radiation and matter domination. We determine it applying the
formula \cite{epole}
\beq w_{\rm rh}=2\frac{\int_{\sgn}^{\sgm} d\sg J(1-
\Vhi/\Vhi(\sgm))^{1/2}}{\int_{\sgn}^{\sgm} d\sg J(1-
\Vhi/\Vhi(\sgm))^{-1/2}}-1,\label{wrh}\eeq\eeqs
where $\sgn=\vev{\sg}$ is the v.e.v of $\sg$ after ETI and $\sgm$
is the amplitude of the $\sg$ oscillations. Motivated by
implementations \cite{univ} of non-thermal leptogenesis, which may
follow ETI, we set $\Trh\simeq10^9~\GeV$ for the reheat
temperature. Finally, $g_{\rm rh}=228.75$ is the energy-density
effective number of degrees of freedom include  corresponding to
the \emph{Minimal SUSY SM} ({\small\sf MSSM}) spectrum.

\subparagraph{\bf (b)}  The amplitude $\As$ of the power spectrum
of the curvature perturbations generated by $\sg$ at  $\ks$ has to
be consistent with data~\cite{plcp}, i.e.,
\begin{equation}  \label{Prob}
A_{\rm s}={\Ve_{\rm I}(\sex)^{3}}/{12\, \pi^2}{\Ve_{\rm
I,\se}(\sex)^2} \simeq2.105\cdot 10^{-9}\,.
\end{equation}

\subparagraph{\bf (c)} The remaining inflationary observables
($\ns$, its running $\as$ and $r$) have to be consistent with the
latest \emph{Planck release 4} ({\sf\small PR4}), \emph{Baryon
Acoustic Oscillations} ({\sf\small BAO}), CMB-lensing and
BICEP/{\it Keck} ({\sf\small BK18}) data \cite{plin,gws}, i.e.,
\begin{equation}  \label{nswmap}
\mbox{\sf
(i)}\>\>\ns=0.965\pm0.009\>\>\>~\mbox{and}\>\>\>~\mbox{\sf
(ii)}\>\>r\leq0.032,
\end{equation}
at 95$\%$ \emph{confidence level} ({\sf c.l.}) -- pertaining to
the $\Lambda$CDM$+r$ framework with $|\as|\ll0.01$. These
observables are estimated through the relations
\beq\label{ns} \mbox{\sf (i)}\>\>\ns=\: 1-6\eph_\star\ +\
2\ith_\star,\>\>\>\mbox{\sf (ii)}\>\> \as
=\frac23\left(4\ith^2-(\ns-1)^2\right)-2\what\xi_\star\>\>\>~
\mbox{and}\>\>\>~\mbox{\sf (iii)}\>\>r=16\eph_\star\,, \eeq
with $\xi={\Ve_{\rm I,\se} \Ve_{\rm I,\se\se\se}/\Ve_{\rm I}^2}$
-- the variables with subscript $\star$ are evaluated at
$\sg=\sgx$.

\subparagraph{\bf (d)}  The effective theory describing ETI has to
remain valid up to a \emph{Ultra Violet} ({\small\sf UV}) cutoff
scale $\Qef\simeq\mP$ to ensure the stability of our inflationary
solutions, i.e.,
\beq \label{uv}\mbox{\sf (i)}\>\> \Vhi(\sgx)^{1/4}\leq\Qef
\>\>\>~\mbox{and}\>\>\>~\mbox{\sf (ii)}\>\>\sgx\leq\Qef.\eeq

Finally, we quantify the tuning needed for the attaintment of
efficient ETI defining
\beq \Dex=1 - \sgx.\label{dex}\eeq
The naturalness of the achievement of ETI increases with $\Dex$.

\subsection{E-Model Inflation}\label{num1}

After imposing \eqs{Nhi}{Prob} the free parameters of
$$\mbox{\dci, ~\ca, ~\cb\ ~~are}~~~(\rs,\mma),~ (N, \rss, \mma)
~~\mbox{and}~~ (N,\rrs,\mmb), $$
respectively. Recall that we use $N=2$ exclusively for \dci.
Fixing $\mma=0.001$ for \dci,  $\mma=0.01$ and $\rss=0.001$ for
\ca\ and $\mmb=0.01$ and $\rrs=0.001$ for \cb, we obtain the
curves plotted and compared to the observational data in
\Fref{fig1}.  We observe that:

\begin{figure}[t]\vspace*{-0.6cm}
\begin{minipage}{75mm}
\includegraphics[height=9cm,angle=-90]{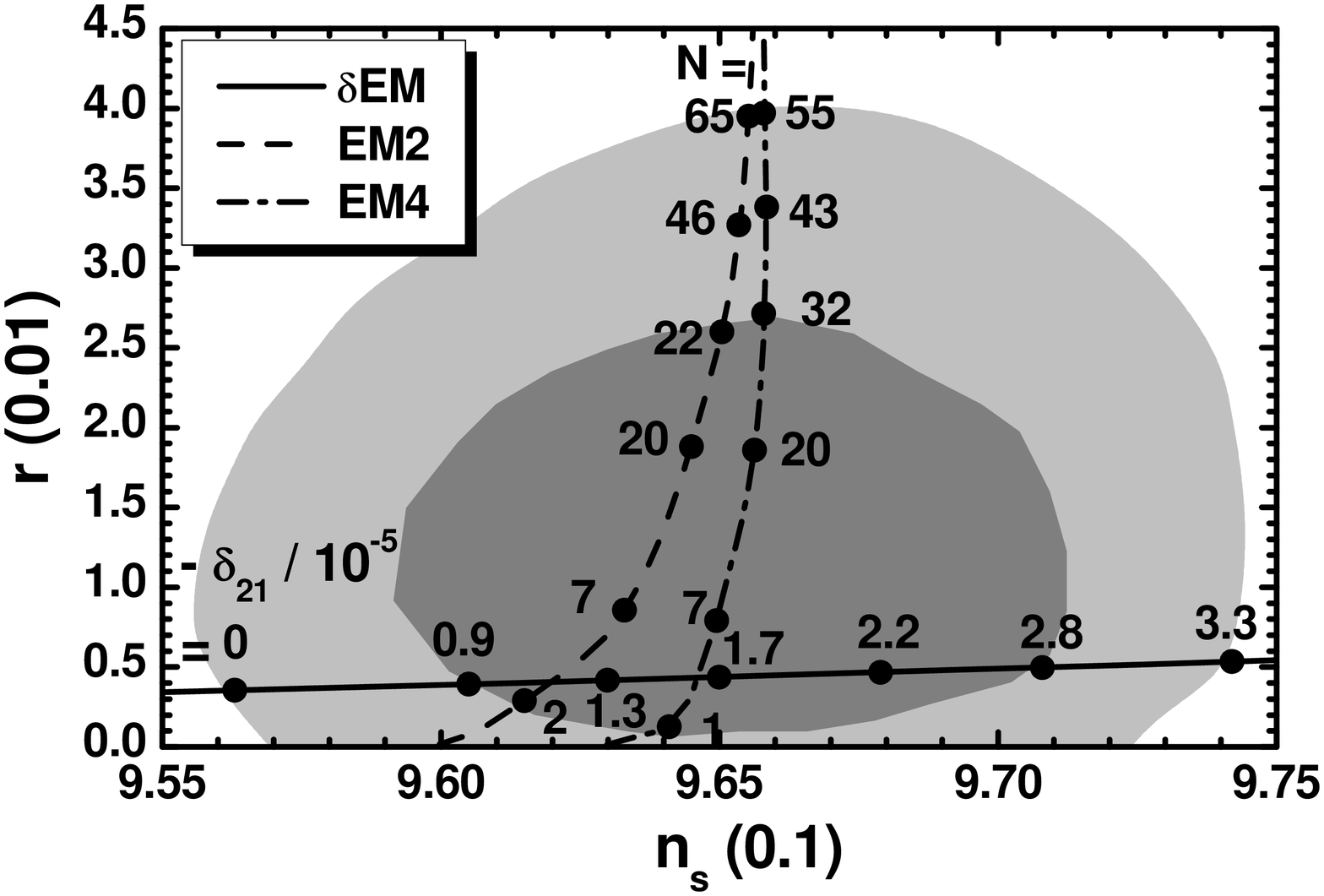}
\end{minipage}
\hfill
\begin{minipage}{95mm}
\begin{center}
\lineup{\small
\begin{tabular}{cccc}\br
{Model:} &{\dci}&{\ca}&{\cb}\cr\mr
${\rs}$ / $\rss$ / $\rrs$&$-1.7\cdot10^{-5}$&{$0.001$}&{$0.001$}\\
$N$&{$2$}&$10$&$10$\cr
$\sgx/0.1$&$9.9$&$9.53$&{$9.84$}\cr
$\Dex (\%)$&$1$&$4.7$&{$2$}\cr
$\sgf/0.1$&{$6.66$}&$3.7$&{$5.6$}\cr\mr
$\wrh$&$-0.24$&$-0.08$&{$0.26$}\cr
$\Ns$&$44.4$&$51.5$&{$55.5$}\cr\mr
$\ld/10^{-5}$&$1.2$&$2.1$&{$1.9$}\cr\mr
$\ns/0.1$&$9.65$&$9.64$&{$9.65$}\cr
$-\as/10^{-4}$&$11.4$ &$6.7$&{$6.2$}\cr
$r/10^{-2}$&$0.44$&$1.3$&{$1.1$}\cr\br
\end{tabular}}
\end{center}
\end{minipage}
\caption{\sl Allowed curves in the $\ns-r$ plane for (i)
\textsf{\small $\delta$EM}, $\mma=0.01$ and various $\rs$'s
indicated on the solid line or (ii) \textsf{\small EM2},
$\mma=0.01$ and $\rss=0.001$ or \textsf{\small EM4}, $\mmb=0.01$
and $\rrs=0.001$ and various $N$'s indicated on the dashed or
dot-dashed line respectively. The marginalized joint $68\%$
[$95\%$] c.l. regions \cite{gws} from PR4, {\sffamily\small BK18},
BAO and lensing data-sets are depicted by the dark [light] shaded
contours. The relevant field values, parameters and observables
corresponding to points shown in the plot are listed in the
Table.} \label{fig1}\end{figure}

\subparagraph{\bf (a)} For \dci\ the resulting $\ns$ and $r$
increase with $|\rs|$ -- see solid line in \Fref{fig1}. From the
considered data we collect the results
\beq \label{resm1} 0\lesssim\rs/10^{-5}\lesssim3.3,
\>\>\>3.5\lesssim {r/10^{-3}}\lesssim5.3\>\>\>
\mbox{and}\>\>\>9\cdot10^{-3}\lesssim\Dex\lesssim0.01. \eeq
In all cases we obtain $\Ns\simeq44$ consistently with \Eref{Nhi}
and the resulting $\wrh\simeq-0.237$ from \Eref{wrh}. Fixing
$\ns=0.965$, we find $\rs=-1.7\cdot 10^{-5}$ and $r=0.0044$ -- see
the leftmost column of the Table in \Fref{fig1}.

\subparagraph{\bf (b)}  For \ca\ and \cb, $\ns$ and $r$ increase
with $N$ and $\Dex$ which increases w.r.t its value in \dci.
Namely, for \ca\ -- see dashed line in \Fref{fig1} -- we obtain
\beqs\beq \label{resm2}
0.96\lesssim\ns\lesssim0.9654,\>\>\>0.1\lesssim N\lesssim
65,\>\>\>0.05\lesssim{\Dex}/{10^{-2}}\lesssim
16.7\>\>\>\mbox{and}\>\>\> 0.0025\lesssim {r}\lesssim0.039\eeq
with $\wrh\simeq-0.05$ and $\Ns\simeq50$. On the other hand, for
\cb\ -- see dot-dashed line in \Fref{fig1} -- we obtain
\beq \label{resm3} 0.963\lesssim\ns\lesssim0.965,\>\>\>0.1\lesssim
N\lesssim 55,\>\>\>0.23\lesssim{\Dex}/{10^{-2}}\lesssim
8.5\>\>\>\mbox{and}\>\>\> 0.0001\lesssim {r}\lesssim0.04\eeq\eeqs
with $\wrh\simeq(0.25-0.39)$ and $\Ns\simeq54-56$. In both
equations above the lower bound on $N$ is just artificial. For
$N=10$, specific values of parameters and observables are arranged
in the rightmost columns of the Table in \Fref{fig1}.

\subsection{T-Model Inflation} \label{num2}

After enforcing \eqs{Nhi}{Prob} -- which yield $\ld$ together with
$\sgx$ -- the free parameters of the models
$$\mbox{\dhi, ~\ha, ~\hb\ ~~are}~~~(\ts,\mmb),~ (N, \tss, \mmb)
~~\mbox{and}~~ (N,\tts,\mmc), $$
respectively. Recall that we use $N=2$ exclusively for \dhi. Also,
we determine $\mmb$ and $\mmc$ demanding that the GUT scale within
MSSM, $\Mgut\simeq2/2.433\times10^{-2}$, is identified with the
value of $M_{BL}$ -- see \Tref{tab1} -- at the vacuum value of
$\sg$, $\vev{\sg}$. We approximately obtain $M_2$ and
$M_4\leq0.001$. By varying the remaining parameters for each model
we obtain the allowed curves in the $\ns-r$ plane-- see
\Fref{fig2}. A comparison with the observational data is also
displayed there.  We observe that:

\begin{figure}[t]\vspace*{-0.6cm}
\begin{minipage}{75mm}
\includegraphics[height=9cm,angle=-90]{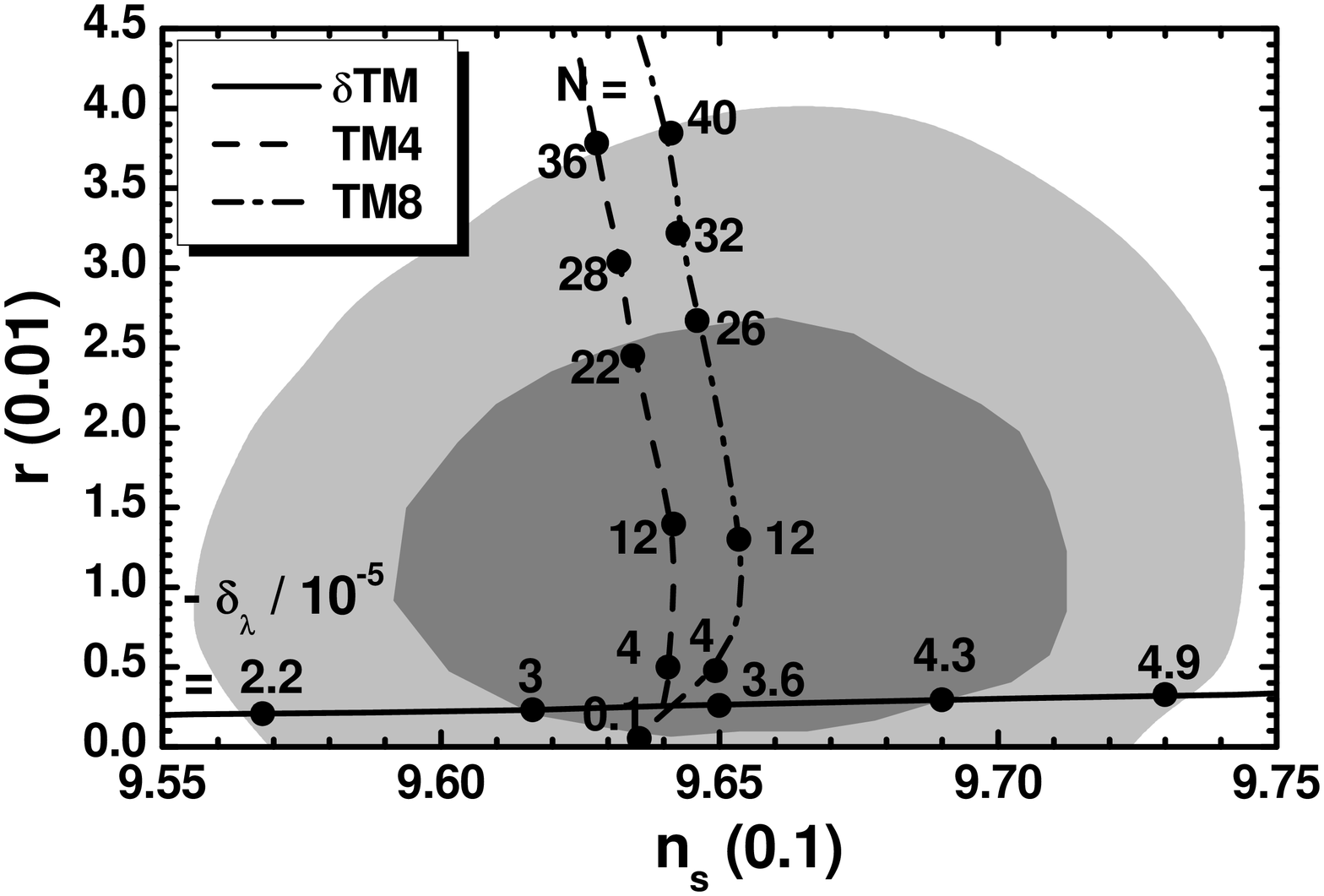}
\end{minipage}
\hfil
\begin{minipage}{95mm}
\begin{center}
\lineup{\small \begin{tabular}{cccc}\br {Model:}
&{\dhi}&{\ha}&{\hb}\cr\mr
${\ts}$ / $\tss$ /
$\tts$&$-3.6\cdot10^{-5}$&{$0.01$}&{$10^{-6}$}\cr
$N$&{$2$}&$12$&$12$\cr\mr
$\sgx/0.1$&$9.9555$&$9.75$&{$9.877$}\cr
$\Dex (\%)$&$0.445$&$2.5$&{$1.23$}\cr
$\sgf/0.1$&{$5.9$}&$3.9$&{$6.5$}\cr\mr
$\wrh$&$0.33$&$0.266$&{$0.58$}\cr
$\Ns$&$55.2$&$56.4$&{$58$}\cr\mr
$\ld/10^{-5}$&$3.6$&$8.6$&{$8.5$}\cr\mr
$\ns/0.1$&$9.65$&$9.64$&{$9.65$}\cr
$-\as/10^{-4}$&$6.6$ &$6.4$&{$5.98$}\cr
$r/10^{-2}$&$0.26$&$1.4$&{$1.3$}\cr\br
\end{tabular}}
\end{center}
\end{minipage}
\caption{\sl Allowed curves in the $\ns-r$ plane fixing
$\mbl=\mgut$ for (i) \textsf{\small $\delta$TM} and various
$\ts$'s indicated on the solid line or (ii) \textsf{\small TM4}
and $\tss=0.01$ or \textsf{\small TM8} and $\tts=10^{-6}$ and
various $N$'s indicated on the dashed and dot-dashed line
respectively. The shaded corridors are identified as in Fig.~1.
The relevant field values, parameters and observables
corresponding to points shown in the plot are listed in the
Table.} \label{fig2}\end{figure}

\subparagraph{\bf (a)} For \dhi\ -- see the solid line in
\Fref{fig2} -- we obtain results similar to those obtained for
\dci\ in \Sref{num1}. Namely, the resulting $\ns$ and $r$ increase
with $|\ts|$ with $\ns$ covering the whole allowed range in
\Eref{nswmap}. From the considered data we collect the results
\beq \label{resh1}2\lesssim-\ts/10^{-5}\lesssim5.5, \>\>\>
2\lesssim
{r/10^{-3}}\lesssim3.6\>\>\>\mbox{and}\>\>\>4\lesssim\Dex/10^{-3}\lesssim4.75.
\eeq
Also, \Eref{Nhi} yields $\Ns\simeq(54.8-55.7)$ and \Eref{wrh}
results to $\wrh\simeq0.3$. Fixing $\ns=0.965$ we find
$\ts=-3.6\cdot 10^{-5}$ and $r=0.0026$ -- see the leftmost column
of the Table in \Fref{fig2}.

\subparagraph{\bf (b)}  For \ha\ and \hb, $\ns$ and $r$ increase
with $N$ and $\Dex$ which is larger than that obtained in \dhi.
More specifically, for \ha\  -- see dashed line in \Fref{fig2} --
we obtain
\beqs\beq \label{resh2}
0.963\lesssim\ns\lesssim0.964,\>\>\>0.1\lesssim N\lesssim
36,\>\>\>0.09\lesssim{\Dex}/{10^{-2}}\lesssim
7.6\>\>\>\mbox{and}\>\>\> 0.0005\lesssim {r}\lesssim0.039\,,\eeq
with $\wrh\simeq0.3$ and $\Ns\simeq56$. On the other hand, for
\hb\ -- see dot-dashed line in \Fref{fig2} -- we obtain
\beq \label{resh3} 0.963\lesssim\ns\lesssim0.965,\>\>\>0.1\lesssim
N\lesssim40,\>\>\>0.45\lesssim{\Dex}/{10^{-2}}\lesssim
3.8\>\>\>\mbox{and}\>\>\> 0.0001\lesssim
{r}\lesssim0.039\,,\eeq\eeqs
with $\wrh\simeq(0.25-0.6)$ and $\Ns\simeq(54.6-60)$. For $N=12$,
specific values of parameters and observables are arranged in the
rightmost columns of the Table in \Fref{fig2}.

%\newpage

\section{Conclusions}\label{con}

We reviewed the implementation of ETI in the context of SUGRA. We
employed as inflaton a gauge singlet or non-singlet superfield for
EMI or TMI respectively. The models are relied on the
superpotentials in \eqs{whi}{whih} which respect an $R$ symmetry
and include, besides inflaton, an inflaton-accompanying field
which facilitates the establishment of ETI. In each model we
singled out three subclasses of models (\dci, \ca\ and \cb) and
(\dhi, \ha\ and \hb). The models \dci\ and \dhi\ are based on the
\Ka s in \sEref{kbas}{a} and (\ref{kbba}{\sf\small a}) whereas
(\ca, \cb) and (\ha, \hb) in those shown in \sEref{kbas}{b} and
(\ref{kbba}{\sf\small b}). Within \dci\ and \dhi\ any
observationally acceptable $\ns$ is attainable by tuning $\rs$ and
$\ts$ respectively to values of the order $10^{-5}$, whereas $r$
is kept at the level of $10^{-3}$ -- see \eqs{resm1}{resh1}. On
the other hand, \ca, \cb, \ha\ and \hb\ avoid any tuning, larger
$r$'s are achievable as $N$ increases beyond $2$, while $\ns$ lies
close to its central observational value -- see \eqs{resm2}{resm3}
for EMI and \eqs{resh2}{resh3} for TMI.

\ack Work supported by the Hellenic Foundation for Research and
Innovation (H.F.R.I.) under the ``First Call for H.F.R.I. Research
Projects to support Faculty members and Researchers and the
procurement of high-cost research equipment grant'' (Project
Number: 2251).

\section*{References}

\def\ijmp#1#2#3{{\sl Int. Jour. Mod. Phys.}
{\bf #1},~#3~(#2)}
\def\plb#1#2#3{{\sl Phys. Lett. B }{\bf #1}, #3 (#2)}
\def\prl#1#2#3{{\sl Phys. Rev. Lett.}
{\bf #1},~#3~(#2)}
\def\rmp#1#2#3{{Rev. Mod. Phys.}
{\bf #1},~#3~(#2)}
\def\prep#1#2#3{{\sl Phys. Rep. }{\bf #1}, #3 (#2)}
\def\prd#1#2#3{{\sl Phys. Rev. D }{\bf #1}, #3 (#2)}
\def\prdn#1#2#3#4{{\sl Phys. Rev. D }{\bf #1}, no. #4, #3 (#2)}
\def\prln#1#2#3#4{{\sl Phys. Rev. Lett. }{\bf #1}, no. #4, #3 (#2)}
\def\npb#1#2#3{{\sl Nucl. Phys. }{\bf B#1}, #3 (#2)}
\def\npps#1#2#3{{Nucl. Phys. B (Proc. Sup.)}
{\bf #1},~#3~(#2)}
\def\mpl#1#2#3{{Mod. Phys. Lett.}
{\bf #1},~#3~(#2)}
\def\jetp#1#2#3{{JETP Lett. }{\bf #1}, #3 (#2)}
\def\app#1#2#3{{Acta Phys. Polon.}
{\bf #1},~#3~(#2)}
\def\ptp#1#2#3{{Prog. Theor. Phys.}
{\bf #1},~#3~(#2)}
\def\n#1#2#3{{Nature }{\bf #1},~#3~(#2)}
\def\apj#1#2#3{{Astrophys. J.}
{\bf #1},~#3~(#2)}
\def\mnras#1#2#3{{MNRAS }{\bf #1},~#3~(#2)}
\def\grg#1#2#3{{Gen. Rel. Grav.}
{\bf #1},~#3~(#2)}
\def\s#1#2#3{{Science }{\bf #1},~#3~(#2)}
\def\ibid#1#2#3{{\it ibid. }{\bf #1},~#3~(#2)}
\def\cpc#1#2#3{{Comput. Phys. Commun.}
{\bf #1},~#3~(#2)}
\def\astp#1#2#3{{Astropart. Phys.}
{\bf #1},~#3~(#2)}
\def\epjc#1#2#3{{Eur. Phys. J. C}
{\bf #1},~#3~(#2)}
\def\jhep#1#2#3{{\sl J. High Energy Phys.}
{\bf #1}, #3 (#2)}
\newcommand\jcap[3]{{\sl J.\ Cosmol.\ Astropart.\ Phys.\ }{\bf #1}, #3 (#2)}
\newcommand\jcapn[4]{{\sl J.\ Cosmol.\ Astropart.\ Phys.\ }{\bf #1}, no. #4, #3 (#2)}
\renewcommand{\arxiv}[1]{\texttt{arXiv:#1}}
\renewcommand{\astroph}[1]{{\tt astro-ph/#1}}
\renewcommand{\hepph}[1]{{\tt hep-ph/#1}}

\end{document}